\documentclass[twocolumn,trackchanges]{aastex6}
\RequirePackage{lineno}
\usepackage{amsmath}
\usepackage{amssymb}
\usepackage{amsthm}
\usepackage{natbib,aasdefs,url,bm}
\usepackage{array}
\usepackage{float}
\usepackage{graphicx}
\usepackage{subfigure}
\usepackage{color}

\newcounter{ichi}
\newcounter{ni}
\newcounter{san}
\newcounter{yon}
\def\be{\begin{equation}}
\def\ee{\end{equation}}
\def\ba{\begin{eqnarray}}
\def\ea{\end{eqnarray}}

\def\aap{\ {A\&A}\ }

\def\apjl{\ {ApJL}\ }
\def\apjs{\ {ApJS}\ }

\shorttitle{Neutrinos from the Brightest Gamma-Ray Burst?}
\shortauthors{Murase et al.}

\begin{document}

\title{Neutrinos from the Brightest Gamma-Ray Burst?}

\author{Kohta Murase\altaffilmark{1,2,3}}
%\affil{Department of Physics, Pennsylvania State University, University Park, PA 16802, USA}

\author{Mainak Mukhopadhyay\altaffilmark{1}}

\author{Ali Kheirandish\altaffilmark{4}}

\author{Shigeo S. Kimura\altaffilmark{5}}

\author{Ke Fang\altaffilmark{6}}

\altaffiltext{1}{Department of Physics; Department of Astronomy \& Astrophysics; Center for Multimessenger Astrophysics, Institute for Gravitation and the Cosmos, The Pennsylvania State University, University Park, PA 16802, USA}
%\altaffiltext{2}{Department of Astronomy \& Astrophysics, The Pennsylvania State University, University Park, PA 16802, USA}
%\altaffiltext{3}{Center for Particle and Gravitational Astrophysics, Institute for Gravitation and the Cosmos, The Pennsylvania State University, University Park, PA 16802, USA}
\altaffiltext{2}{School of Natural Sciences, Institute for Advanced Study, Princeton, NJ 08540, USA}
\altaffiltext{3}{Center for Gravitational Physics and Quantum Information, Yukawa Institute for Theoretical Physics, Kyoto University, Kyoto, Kyoto 606-8502, Japan}
\altaffiltext{4}{Department of Physics \& Astronomy; Nevada Center for Astrophysics, University of Nevada, Las Vegas, NV 89154, USA}
%\altaffiltext{7}{Nevada Center for Astrophysics, University of Nevada, Las Vegas, NV, 89154}
\altaffiltext{5}{Frontier Research Institute for Interdisciplinary Sciences; Astronomical Institute, Graduate School of Science, Tohoku University, Sendai 980-8578, Japan}
%\altaffiltext{9}{Astronomical Institute, Graduate School of Science, Tohoku University, Sendai 980-8578, Japan}
\altaffiltext{6}{Department of Physics,  Wisconsin IceCube Particle Astrophysics Center, University of Wisconsin, Madison, WI 53706, USA}

\begin{abstract}
We discuss implications that can be obtained by searches for neutrinos from the brightest gamma-ray burst, GRB 221009A. 
We derive constraints on GRB model parameters such as the cosmic-ray loading factor and dissipation radius, taking into account both neutrino spectra and effective areas. The results are strong enough to constrain proton acceleration near the photosphere, and we find that the single burst limits are comparable to those from stacking analysis. 
Quasithermal neutrinos from subphotospheres and ultrahigh-energy neutrinos from external shocks are not yet constrained. We show that GeV-TeV neutrinos originating from neutron collisions are detectable, and urge dedicated analysis on these neutrinos with DeepCore and IceCube as well as ORCA and KM3NeT. 
\end{abstract}

%\keywords{neutrinos}

\section{Introduction}
Gamma-ray bursts (GRBs) are the most powerful explosive phenomena in the Universe, which have been extensively discussed as ultrahigh-energy (UHE) cosmic-ray (CR) accelerators~\citep{Waxman:1995vg,Vietri:1995hs}. Accompanied high-energy neutrinos have been searched for, but no detection has been reported so far~\citep{Abbasi:2012zw,IceCube:2017amx,ANTARES:2020vzs}. Canonical high-luminosity GRBs cannot make a major contribution to the all-sky neutrino flux measured in IceCube, and optimistic cases have been ruled out. The hypothesis that UHE CRs come from GRBs has not yet been excluded, and various possibilities of high-energy CR and neutrino production in GRBs have been investigated~\citep[see reviews][and references therein]{Meszaros:2015krr,Kimura:2022zyg}.    

On October 9, 2022, an extraordinarily bright burst, GRB 221009A was reported. This burst was reported by {\it Swift}-BAT as an unknown-type transient~\citep{221009ABAT}, and it had triggered {\it Fermi}-GBM about one hour before the BAT trigger time \citep{221009AGBM}. The burst showed an initial pulse of $\sim 10$~s, followed by the main burst beginning at $\sim 180$~s after the GBM trigger time. The preliminary estimate of the gamma-ray energy fluence reported by Konus-Wind is $\sim5\times{10}^{-2}~\rm erg~cm^{-2}$~\citep{221009KW}. Afterglow emission has been observed at different wavelengths, and optical follow-up observations revealed that the redshift of this GRB is $z=0.15$~\citep{221009VLT}, which suggests that the isotropic-equivalent gamma-ray energy is ${\mathcal E}_{\gamma}^{\rm iso}\sim3\times10^{54}$~erg. 
The detection of high-energy gamma-rays at $\sim200-600$~s after the GBM trigger time was reported by {\it Fermi}-LAT. The highest-energy photon of 99 GeV was detected by LAT at 240~s after the trigger~\citep{221009AFermi}. The Large High Altitude Air Shower Observatory observed more than 5000 very-high-energy photons in the TeV range, and even $\gtrsim10$~TeV photons were detected~\citep{221009LHAASO}. 

In this work, we demonstrate how observations of neutrinos from the brightest GRB can be used for learning about models of GRB neutrino emission. We focus on neutrinos emitted during the prompt phase, and consider not only nonthermal neutrinos accompanied by CR acceleration but also quasithermal neutrinos produced by inelastic collisions with neutrons. We use $Q_x/Q=10^{x}$ in CGS units and assume cosmological parameters with $\Omega_m=0.3$, $\Omega_\Lambda=0.7$ and $h=0.7$.

\section{Nonthermal Emission}\label{sec:nonthermal}
\subsection{Gamma-ray constraints}
The detection of high-energy gamma-rays can be used for placing a lower limit on the bulk Lorentz factor $\Gamma$~\citep[e.g.,][]{Lithwick:2000kh} and/or the dissipation radius $r_{\rm diss}$~\citep[e.g.,][]{Murase:2007ya,Gupta:2007mx,Zhang:2009aca}. 
The detection of a $\sim100$~GeV photon at $\sim240$~s after the trigger~\citep{221009AFermi} suggests that the emission region has to be transparent to $\gamma\gamma\rightarrow e^+e^-$. The two-photon annihilation optical depth for a gamma-ray with energy $\varepsilon_\gamma$ is 
\begin{equation}
\tau_{\gamma\gamma}\simeq150\frac{\eta_{\gamma\gamma,-1}L_{\gamma,52.5}}{r_{\rm diss,14}\Gamma_{2.5}^2 \varepsilon_{\rm MeV}^{b}} 
\left\{
\begin{array}{rl} 
{(\varepsilon_\gamma/\varepsilon_\gamma^b)}^{\beta-1} & \mbox{($\varepsilon_\gamma < \varepsilon_{\gamma}^{b}$)}\\
{(\varepsilon_{\gamma}/\varepsilon_{\gamma}^{b})}^{\alpha-1} & \mbox{($\varepsilon_\gamma^{b} < \varepsilon_\gamma$)} 
\end{array} \right., \label{eq:gg}
\end{equation}
where $\eta_{\gamma\gamma}\sim0.1$ is a spectrum-dependent coefficient~\citep{sve87}, $L_{\gamma}$ is the isotropic-equivalent gamma-ray luminosity during the main brightest episode, where we take $10^{52.5}~{\rm erg}~{\rm s}^{-1}$ in the Konus-Wind band (so that the band correction is included), $\varepsilon^{b}\equiv1~{\rm MeV}~\varepsilon_{\rm MeV}^{b}$ is the photon break energy in the GRB frame, $\alpha$ and $\beta$ are low- and high-energy photon indices, respectively. The typical energy of high-energy gamma-rays interacting with target photons at a break energy is $E_\gamma^b=\varepsilon_{\gamma}^{b}/(1+z)\approx \Gamma^2m_e^2c^4/\varepsilon^b/(1+z)\simeq 23~{\rm GeV}~\Gamma_{2.5}^2{(\varepsilon_{\rm MeV}^{b})}^{-1}$ for GRB 221009A.

Requiring $\tau_{\gamma\gamma}(E_\gamma=100~\rm GeV)<1$, with $\varepsilon^{b}\sim1~{\rm MeV}$, $\alpha\sim1.0$ and $\beta\sim2.6$~\citep{221009KW}, the dissipation radius can be constrained as
\begin{equation}
r_{\rm diss}\gtrsim1.5\times{10}^{16}~{\rm cm}~L_{\gamma,52.5}\Gamma_{2.5}^{-2}
\left\{\begin{array}{rl} 
1\,\,\,\,\,\,\,\,\,\,\,\,\,\, & \mbox{($\Gamma\lesssim700$)}\\
{(5\Gamma_{2.5}^{-2})}^{1.6} & \mbox{($\Gamma\gtrsim700$)} 
\end{array} \right.. \label{eq:gg}
\end{equation}
We also obtain $\Gamma\gtrsim770~{(L_{\gamma,52.5}{\delta t}_{-2}^{-1})}^{5/36}$ with $r_{\rm diss}\approx2\Gamma^2 c\delta t/(1+z)$ that is expected in the internal shock scenario (where $\delta t$ is the variability time scale), although this constraint can be relaxed in multi-zone models~\citep{Aoi:2009ty}. High-energy gamma-rays with $\varepsilon_\gamma \gg 1$~GeV are unlikely to be produced near the photosphere, as has been argued for some of the past bright GRBs~\citep[e.g.,][]{Zhang:2009aca}.

\subsection{Neutrino constraints}
%
%%%%%%%%%%%%%%%%%%%%%%%%%%%%%%%%%%
\begin{figure*}[th]
\begin{center}
\includegraphics[width=0.32\linewidth]{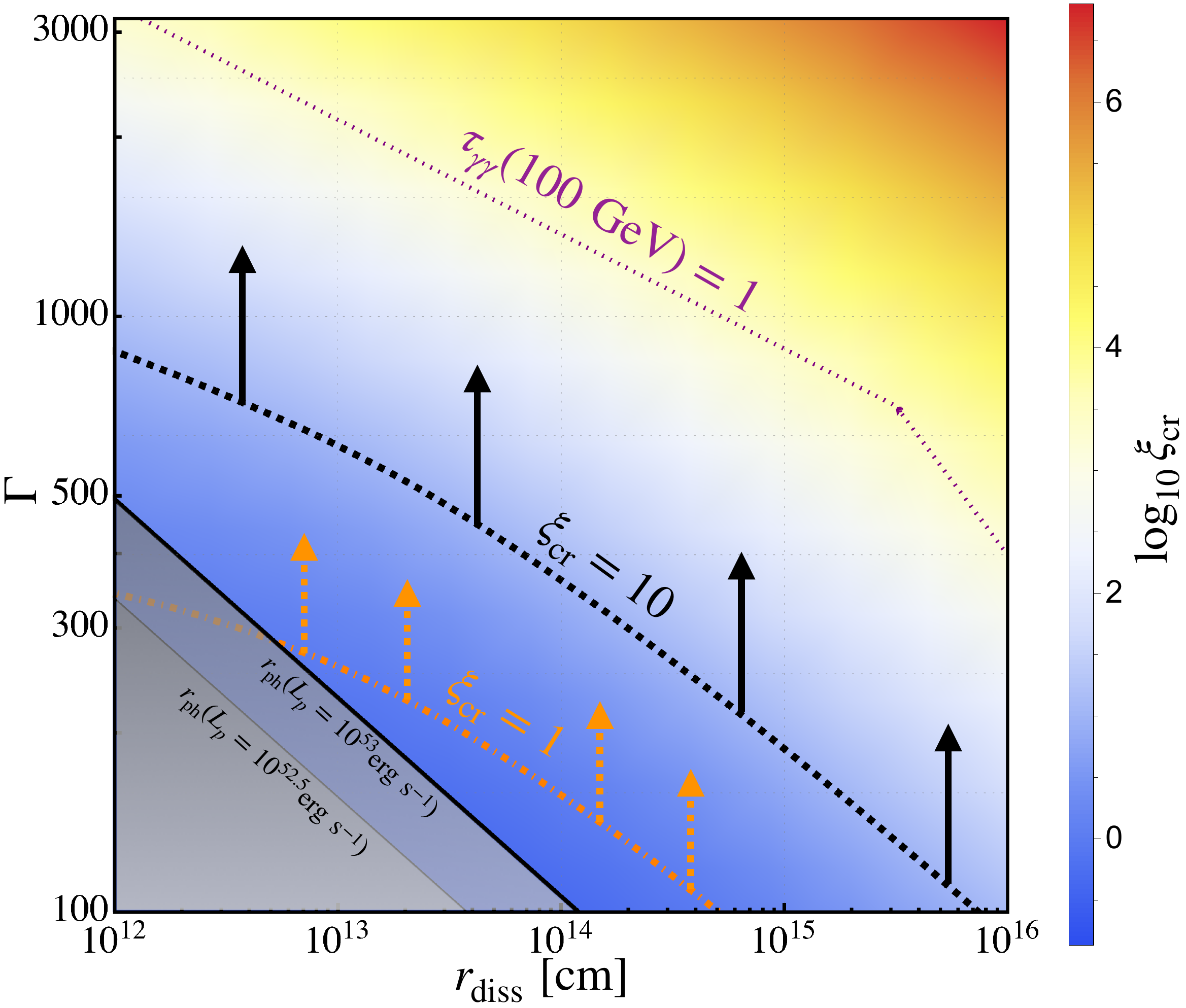}
\includegraphics[width=0.32\linewidth]{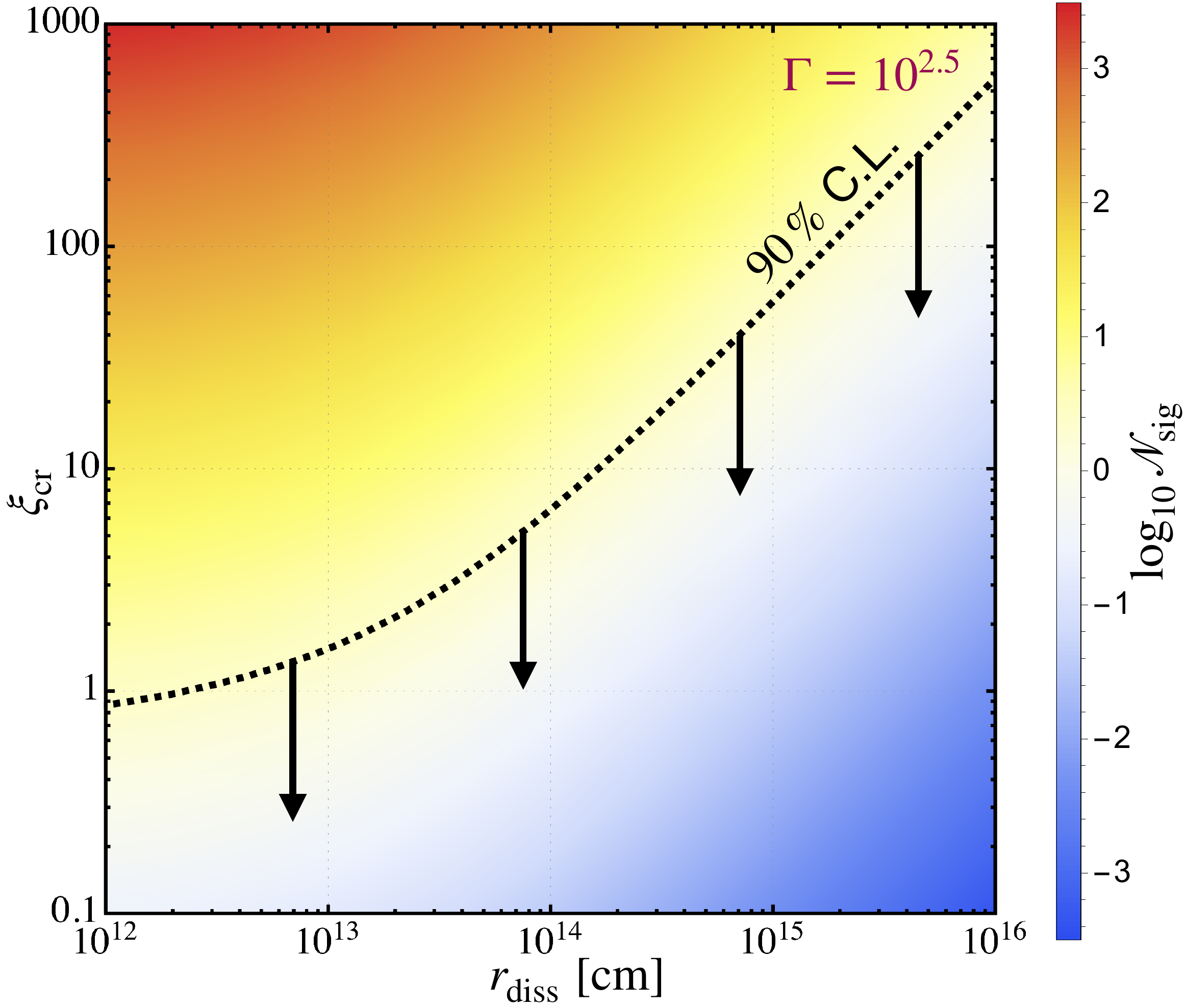}
\includegraphics[width=0.32\linewidth]{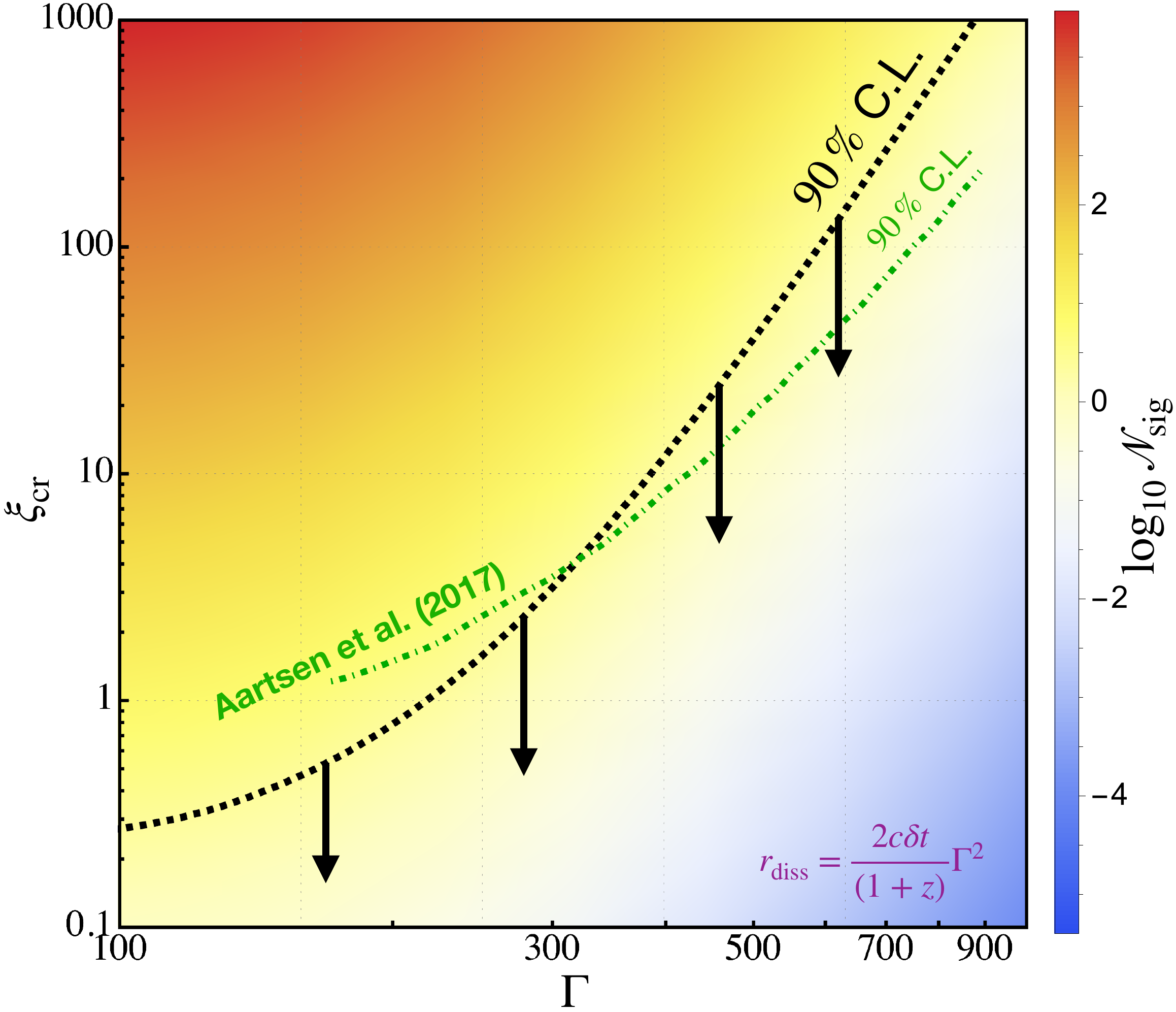}
\caption{
(Left): Constraints on $\Gamma$ as a function of $r_{\rm diss}$ for different values of the CR loading factor $\xi_{\rm cr}$. The region below $r_{\rm ph}$ is not considered for nonthermal neutrino production. 
(Middle): Constraints on $\xi_{\rm cr}$ as a function of $r_{\rm diss}$ for a given Lorentz factor of $\Gamma=10^{2.5}$, where color scale represents the number of signal events ${\mathcal N}_{\rm sig}$. 
(Right): Constraints on $\xi_{\rm cr}$ as a function of $\Gamma$, where the internal shock model is assumed with $\delta t=0.01$~s. The IceCube stacking limit at $90\%$ C.L. \citep{IceCube:2017amx} for this model is also shown.
}
\label{fig:ntnu}
%\vspace{-1.\baselineskip}
\end{center}
\end{figure*}
%%%%%%%%%%%%%%%%%%%%%%%%%%%%%%%%%%

%Let us consider a generic source with comoving size $r/\Gamma$ (where $r$ is the emission radius and $\Gamma$ is the bulk Lorentz factor of the source).  We assume the presence of target photons with $\varepsilon'_t\approx\varepsilon_t/\Gamma$ and spectrum $n_{\varepsilon'_t}$.  
%For $n_{\varepsilon'_t}\propto{\varepsilon'}_t^{-\alpha}$ with $\alpha>1$, which is valid in most nonthermal objects, meson production is dominated by the $\Delta$-resonance and direct pion production.  Its efficiency $f_{p\gamma}$ is given by
%\begin{equation}\label{eq:fpgamma}
%f_{p \gamma}(\varepsilon_p)\approx\eta_{p\gamma}(\alpha)\hat{\sigma}_{p\gamma}(r/\Gamma)(\varepsilon'_t n_{\varepsilon'_t})|_{\varepsilon'_t=0.5m_pc^2\bar{\varepsilon}_{\Delta}/\varepsilon'_p}\,,
%\end{equation}
%where $\hat{\sigma}_{p\gamma}\sim0.7\times{10}^{-28}~{\rm cm}^2$ is the attenuation cross section (the product of the inelasticity and cross section~\cite{Dermer:2010hy,Murase:2008mr,Dermer:2014vaa}), $\eta_{p\gamma}(\alpha)\approx2/(1+\alpha)$, and $\bar{\varepsilon}_\Delta\sim0.3$~GeV.  The energy of protons that typically interact with photons with cosmic reference frame energy $\varepsilon_t$ is $\varepsilon_p\approx20\varepsilon_\nu\approx0.5\Gamma^2m_pc^2\bar{\varepsilon}_\Delta{\varepsilon_t}^{-1}$, leading to $\varepsilon_t\sim20~{\rm keV}~{(\Gamma/10)}^2{(\varepsilon_\nu/{\rm 30~TeV})}^{-1}$.  Thus, the IceCube data imply sources with {\it x-ray or MeV $\gamma$-ray counterparts}.  

If the high-energy CRs are accelerated during the prompt phase, they should interact with GRB photons via the photomeson production process \citep{Waxman:1997ti}, leading to a flux of high-energy neutrinos via decay processes like $\pi^+\to\mu^+\nu_\mu\to\nu_\mu\bar{\nu}_\mu\nu_ee^+$. The effective $p\gamma$ optical depth is~\citep[e.g.,][]{Waxman:1997ti,Murase:2006mm}
\begin{equation}
f_{p\gamma} \simeq 0.15 \frac{\eta_{p\gamma}L_{\gamma,52.5}}{r_{\rm diss,14} \Gamma _{2.5}^2
\varepsilon_{\rm MeV}^{b}} \left\{ \begin{array}{lr} 
{(\varepsilon_p/\varepsilon_p^b)}^{\beta-1} & \mbox{($\varepsilon_p < \varepsilon_{p}^{b}$)}\\
{(\varepsilon_{p}/\varepsilon_{p}^{b})}^{\alpha-1} & \mbox{($\varepsilon_p^{b} < \varepsilon_p$)} 
\end{array} \right., 
\label{eq:pg}
\end{equation}
where $\varepsilon_{p}^{b}\approx0.5 \bar{\varepsilon}_{\Delta}m_pc^2 \Gamma^2/\varepsilon^{b}$ is the proton break energy in the GRB frame, and $\bar{\varepsilon}_{\Delta}\sim0.3$~GeV. Here $\eta_{p\gamma}$ is a correction factor that is $\sim (2-3)$ for $\alpha\sim1$ due to the effects of multipion production and high inelasticity~\citep{Murase:2005hy}. The resulting typical neutrino energy is $E_\nu^b\approx0.05\varepsilon_{p}^{b}/(1+z)\sim0.6~{\rm PeV}~\Gamma_{2.5}^2{(\varepsilon_{\rm MeV}^{b})}^{-1}$. 

By introducing the CR loading factor $\xi_{\rm cr}\equiv{\mathcal E}_{\rm cr}^{\rm iso}/{\mathcal E}_{\gamma}^{\rm iso}$ ~\citep{Murase:2005hy}, the neutrino fluence
\begin{eqnarray}
E_\nu^2 \phi_{\nu_\mu}&\approx&\frac{1}{8}\frac{(1+z)}{4 \pi d_L^2}{\rm min}[1,f_{p\gamma}]\xi_{\rm cr}\frac{{\mathcal E}_\gamma^{\rm iso}}{\mathcal R_{\rm cr}}\nonumber\\
&\sim&4\times{10}^{-3}~{\rm erg}~{\rm cm}^{-2}~{\rm min}[1,f_{p\gamma}]\xi_{\rm cr,1}{\mathcal E}_{\gamma,54.5}^{\rm iso}{\mathcal R_{\rm cr,1.2}^{-1}},\,\,\,\,\,\,
\end{eqnarray}
where $1/8$ comes from the facts that the $\pi^\pm/\pi^0$ ratio is $\sim1$ in $p\gamma$ interactions due to the contribution from direct production 
%near the $\Delta$ resonance 
and each flavor of neutrinos after the mixing carries $\sim1/4$ of the pion energy in the decay chain. Also, ${\mathcal R}_{\rm cr}$ is a spectrum dependent factor that converts the bolometric CR energy to the differential CR energy, which is ${\mathcal R}_{\rm cr}\sim15-20$ for a CR spectral index of $s_{\rm cr}=2.0$ depending on the CR maximum energy. 

Non-detection of neutrinos from GRB 221009A was reported by \cite{221009AIC}, which gives $E_\nu^2 \phi_{\nu_\mu}\leq3.9\times10^{-2}~{\rm GeV}~{\rm cm}^{-2}$ at 90\% C.L. for an $E_\nu^{-2}$ spectrum. This naively infers 
\begin{equation}
{\rm min}[\xi_{\rm cr,-1},f_{p\gamma,-2}\xi_{\rm cr,1}]\lesssim2.
\end{equation}
However, this constraint is optimistic and it should not be used in general cases. Since GRB neutrino spectra are not described by a single power law, it is significantly relaxed when $E_\nu^b$ is higher than $10-100$~TeV, the regime in which IceCube is the most sensitive \citep{IceCube:2020mzw}. 

Because the dissipation radius of prompt emission is not well-known and under debate, it is often more useful to treat $r_{\rm diss}$ as an uncertain parameter ~\citep{Murase:2008mr,Zhang:2012qy}. In Figure~\ref{fig:ntnu} left and middle, we present constraints in the $r_{\rm diss}-\Gamma$ plane~\citep[see also][for GRB 130427A]{Gao:2013fra} and $r_{\rm diss}-\xi_{\rm cr}$ plane, respectively. The neutrino spectra are calculated using the prescription in \cite{He:2012tq} and \cite{Kimura:2017kan}, assuming $\xi_B=1$ for magnetic fields~\citep{Murase:2005hy}.
We adopt $\varepsilon^{b}=1.2~{\rm MeV}$, $\alpha=1.1$ and $\beta=2.6$~\citep{221009KW}, which is sufficient for the purpose of this work to demonstrate the constraints and to encourage further searches with detailed information on time-dependent spectra. 
We use the point source effective area\footnote{In general the GFU effective area \citep{IceCube:2016cqr} should be used for real time follow-ups. But the publicly available data do not have a sufficiently fine binning in the zenith angle.} \citep{IceCube:2021xar} at declination, $\delta\approx 19.8^\circ$. The convolution of the neutrino flux and the effective area over a relevant energy range ($200~{\rm GeV}\leq E_\nu \leq 10^9~{\rm GeV}$) gives the number of signal events, ${\mathcal N}_{\rm sig}$. Our limit for an $E_\nu^{-2}$ spectrum also agrees with the IceCube limit \citep{221009AIC}. The results on ${\mathcal N}_{\rm sig}$ are not strongly affected by $\xi_B$. This is because the signal mainly comes from neutrinos around $E_\nu^b$, whereas $\xi_B$ is important for the neutrino flux suppression that occurs at much higher energies at $\sim10-1000$~PeV~\citep{Murase:2005hy}.

Remarkably, we obtain strong constraints on particle acceleration near the photosphere at $r_{\rm ph}\simeq3.8\times{10}^{12}~{\rm cm}~\zeta_e L_{p,53}\Gamma_{2.5}^{-3}$
%\begin{equation}
%r_{\rm ph}\simeq1.2\times{10}^{13}~{\rm cm}~\zeta_e L_{p,53.5}\Gamma_{2.5}^{-3},    
%\end{equation}
in the limit that the coasting occurs under $r_{\rm ph}$. Here $L_p$ is the proton luminosity and $\zeta_e$ is the number ratio of electrons and positrons to protons. 
From Figure~\ref{fig:ntnu} left and middle, we obtain $\xi_{\rm cr}\lesssim1$ for $\Gamma\lesssim300$, which excludes the benchmark case of the baryonic photospheric scenario ($\xi_{\rm cr}=1$ and $\zeta_e=1$), although the limits can be relaxed with larger values of $\Gamma$.   
Note that these constraints on the baryonic photospheric scenario are largely insensitive to uncertainty in $L_\gamma$ because of $f_{p\gamma}\sim20 (L_{\gamma,52.5}/L_{p,53})(\Gamma_{2.5}/\varepsilon_{\rm MeV}^b)\tau_T\gg1$ near the photosphere~\citep{Murase:2008sp}, where $\tau_T$ is the Thomson optical depth. Our results are conservative in the sense that we do not include $pp$ collisions that are relevant in the TeV range~\citep{Murase:2008sp,Wang:2008zm}.

On the other hand, IceCube's non-detection is consistent with outer-zone (i.e., large $r_{\rm diss}$) models. For example, we obtain $r_{\rm diss}\gtrsim(2-20)\times{10}^{14}$~cm for $\Gamma\sim300$ and $\xi_{\rm cr}\sim10-100$. Such parameter space is favored by the scenario where UHE CRs are nuclei rather than protons~\citep[see Figure 8 of][]{Murase:2008mr} and some of the magnetic reconnection models~\citep[e.g.,][]{Zhang:2012qy,Pitik:2021xhb}. This also rules out the neutron escape scenario for UHE CRs~\citep{Ahlers:2011jj}. 
We also note that low efficiencies of the photomeson production process are also consistent with the detection of a $\sim100$~GeV photon. From Eqs.~(\ref{eq:gg}) and (\ref{eq:pg}) we obtain
\begin{equation}
f_{p\gamma}\lesssim3\times{10}^{-3}~\eta_{p\gamma,0.5}
\left\{\begin{array}{lr} 
1 & \mbox{($\Gamma\lesssim700$)}\\
{(\Gamma_{2.5}^{2}/5)}^{1.6} & \mbox{($\Gamma\gtrsim700$)} 
\end{array} \right..
\label{eq:pg-gg}
\end{equation}
This limit does not depend on $L_\gamma$ and $r_{\rm diss}$, and it is applicable to all proton energies given $\alpha\sim1$. Although Eq.~(\ref{eq:pg-gg}) is robust as long as neutrinos and gamma-rays are co-produced, it is worthwhile to note that their emission regions may be different. For example, in the photospheric scenario, sub-TeV gamma-rays are unlikely to escape and hence should come from large dissipation radii, e.g., at the external reverse shock. 

Prompt GRB neutrinos have been best studied in the context of the internal shock scenario, and the UHE CR hypothesis requires $\xi_{\rm cr}\sim10-100$~\citep{Murase:2008mr,Biehl:2017qen}. The constraints with the assumption of $r_{\rm diss}\approx2\Gamma^2 c\delta t/(1+z)$ are presented in Figure~\ref{fig:ntnu} right. Here, for the purpose of the comparison with the IceCube result~\citep{IceCube:2017amx}, we use $\delta t=0.01$~s, although the chosen value is subject to both observational and model uncertainties~\citep[e.g.,][]{Murase:2005hy,Murase:2008mr,Zhang:2012qy}.

We find $\xi_{\rm cr}\lesssim3$ for $\Gamma=300$. This implies that for a benchmark Lorentz factor of $\Gamma=300$ that is often used in the literature~\citep[e.g.,][]{IceCube:2017amx}, the case motivated by the UHE CR hypothesis may be excluded, where the constraint given in Eq~(\ref{eq:gg}) should be alleviated if neutrinos and the highest-energy gamma-rays come from different regions~\citep[e.g.,][]{Bustamante:2014oka,Zhang:2022lff}. Alternatively, GRB internal shocks are still viable for UHE CR acceleration if the Lorentz factor is high enough to lead to large $r_{\rm diss}\gtrsim(2-20)\times{10}^{14}$~cm, as used in \cite{Murase:2008mr}. 
Interestingly, our new limit shown in Figure~\ref{fig:ntnu} is comparable to the IceCube stacking limit \citep{IceCube:2017amx}. Our results are useful because the latter is subject to systematic uncertainties coming from the aggregation of many bursts. One single burst provides complementary constraints, and support that canonical high-luminosity GRBs contribute less than $\sim1$\% of the all-sky neutrino flux.

%\subsection{All-Sky Neutrino Flux}
%
%The all-sky neutrino flux for the sum of all flavors can be estimated to be
%\begin{eqnarray}
%E_\nu^2\Phi_\nu&\sim&\frac{c}{4\pi H_0}\frac{3}{8}{\rm min}[1,f_{p\gamma}]\frac{{\mathcal E}_\gamma^{\rm iso}}{\mathcal R_{\rm cr}}\rho \xi_z\\
%&\sim&4\times{10}^{-8}~{\rm GeV}~{\rm cm}^{-2}~{\rm s}^{-1}~{\rm sr}^{-1}~\xi_{\rm cr,1}\nonumber\\
%&\times&{\rm min}[1,f_{p\gamma}]({\mathcal E}_{\gamma}^{\rm iso}\rho/{10}^{53}~{\rm erg}~{\rm Gpc}^{-3}~{\rm yr}^{-1})(\xi_z/3),\nonumber
%\end{eqnarray}
%where $\xi_z$ is the evolution factor~\citep{Waxman:1997ti} and $\rho\sim1~{\rm Gpc}^{-3}~{\rm yr}^{-1}$ is the apparent GRB rate density at $z=0$.  

\section{Quasithermal Emission}\label{sec2}
Subphotospheric neutrino production at $\tau_{\rm T}\gtrsim1$ is efficient if CRs exist. However, CR acceleration at radiation-mediated shocks is inefficient, and the detection of nonthermal neutrinos from deep subphotospheres is unlikely for canonical high-luminosity GRBs~\citep{Murase:2013ffa}. 
However, high-energy neutrinos can still be produced without relying on CR acceleration. Neutrons can provide neutrinos through inelastic collisions between bulk flows or neutron diffusion \citep{Meszaros:2000fs}, without involving collisionless shocks or magnetic reconnections. Such ``quasithermal'' neutrinos are naturally produced during neutron decoupling \citep{Bahcall:2000sa} and/or by internal collisions between neutron-loaded outflows \citep{Murase:2013hh,Bartos:2013hf,Zegarelli:2021vuf}.

%%%%%%%%%%%%%%%%%%%%%%%%%%%%%%%%%%
\begin{figure*}[th]
\begin{center}
\includegraphics[width=0.48\linewidth]{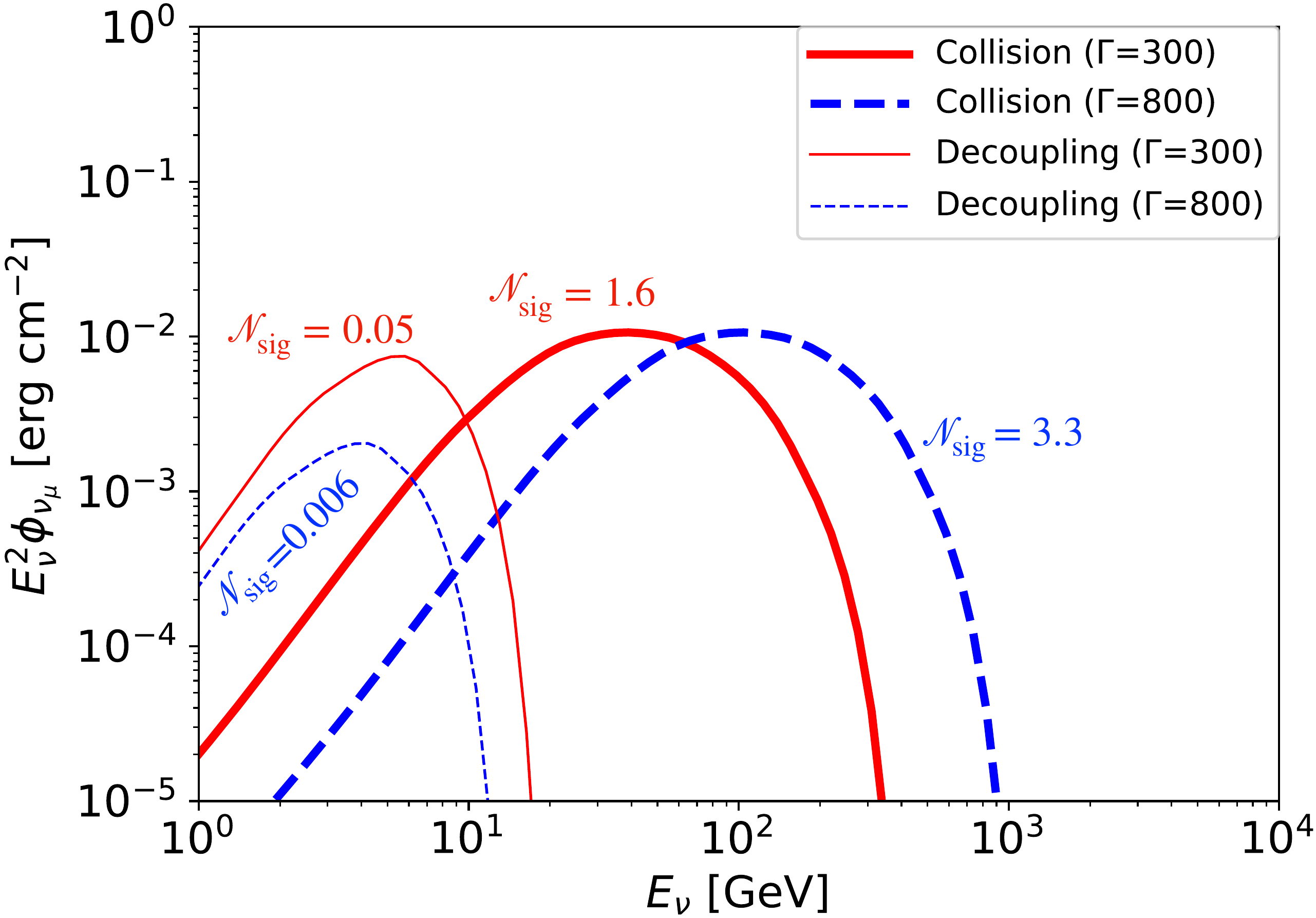}
\includegraphics[width=0.49\linewidth]{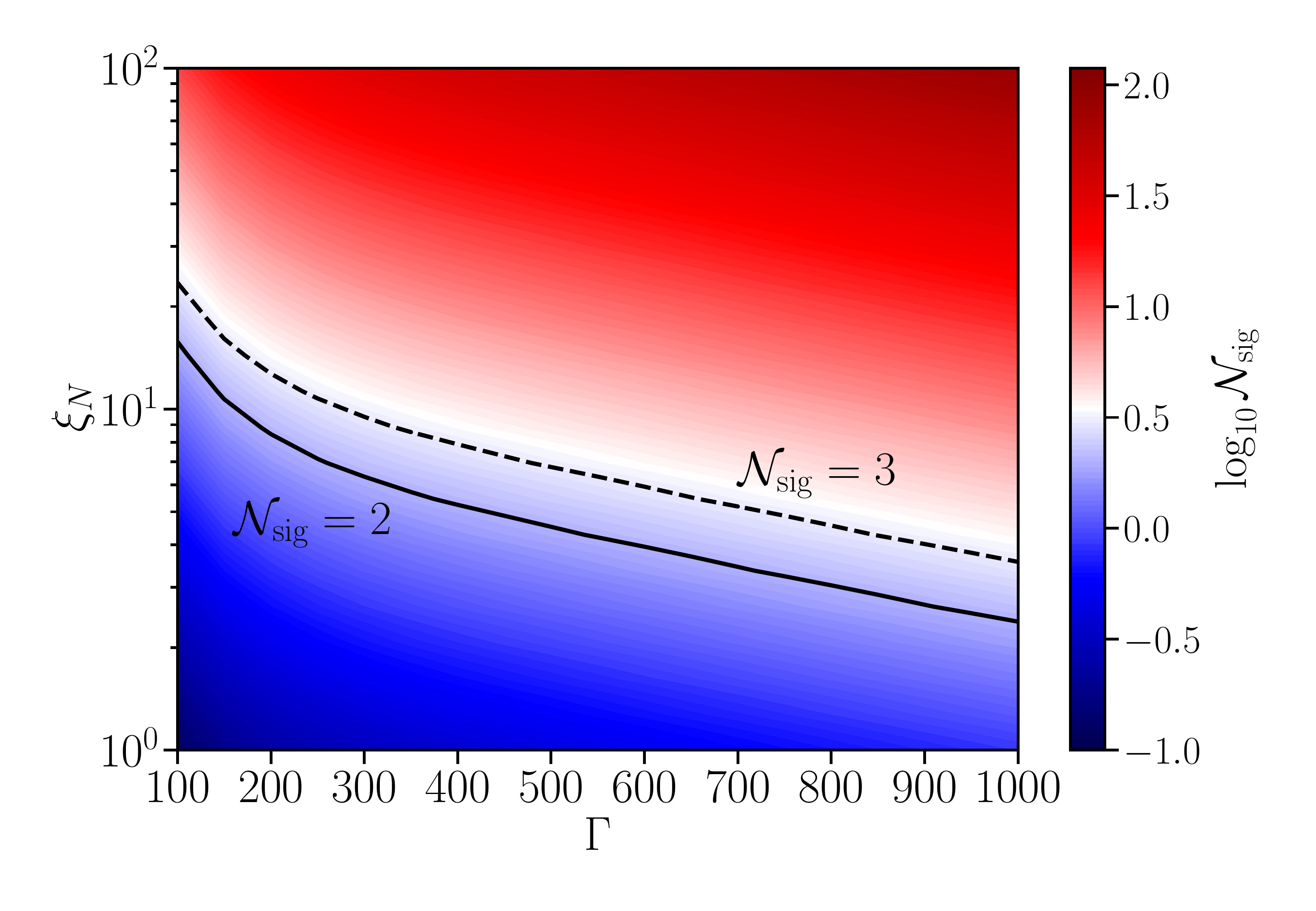}
\caption{Left: Energy fluences of quasithermal $\nu_\mu$ from GRB 221009A for both collision and decoupling scenarios, where $\xi_N=5$ and ${\mathcal E}_{\gamma}^{\rm iso}=10^{54.5}$~erg are used. Right: Expected number of signal events, ${\mathcal N}_{\rm sig}$, in DeepCore+IceCube as a function of $\xi_N$ and $\Gamma$. The solid and dashed lines show the parameter sets that lead to doublet and triplet events, respectively.}
\label{fig:qtnu}
%\vspace{-1.\baselineskip}
\end{center}
\end{figure*}
%%%%%%%%%%%%%%%%%%%%%%%%%%%%%%%%%%

\subsection{Neutrinos from neutron decoupling}
Recent studies have shown that a GRB jet is collimated during its propagation inside a star~\citep[e.g.,][]{Bromberg:2011fg,Mizuta:2013yma,Hamidani:2020krf,Gottlieb:2022tkb}. The jet material becomes hot and the post-collimation density is so high that $\tau_{\rm T}\gg\tau_{np}\gg1$, in which protons and neutrons are coupled. After the breakout, the hot jet material may expand with $\Gamma(r)\approx \Gamma_*(r/R_*)$ like a fireball, where $\Gamma_*\approx1/\theta_j=10~\theta_{j,-1}^{-1}$ is the Lorentz factor at the breakout. By equating the $np$ collision time $t'_{np}\approx1/(n'_p\sigma_{np}c)$ (where $\sigma_{np}\approx3\times{10}^{-26}~{\rm cm}^2$ is the approximate $np$ cross section, $n'_p\approx L_p/(4\pi r^2\Gamma\Gamma_{\rm max}m_p c^3)$ is the proton density and $\Gamma_{\rm max}$ is the maximum Lorentz factor) and the expansion time $t'_{\rm dyn}\approx r/(\Gamma c)$, the decoupling radius is estimated to be $r_{\rm dec}\approx8.7\times{10}^{11}~{\rm cm}~L_{p,53}^{1/3}R_{*,11}^{2/3}\Gamma_{*,1}^{-2/3}\Gamma_{\rm max,2.9}^{-1/3}$, at which the Lorentz factor becomes
\begin{equation}
\Gamma_{n,\rm dec}\approx65~L_{p,53}^{1/3}R_{*,11}^{-1/3}\Gamma_{*,1}^{1/3}\Gamma_{\rm max,2.9}^{-1/3},
\label{eq:ndec}
\end{equation}
where a numerical factor of $3/4$ is taken into account. Note that the decoupling during the acceleration occurs if $\Gamma_{\rm max}$ is larger than the critical value,
\begin{equation}
\Gamma_{np}\approx{\left(\frac{\sigma_{np}L_p\Gamma_*}{4\pi m_pc^3 R_*}\right)}^{1/4}.
\end{equation} 

Radiative acceleration is fast and the flow is accelerated relativistically, in which inelastic $np$ collisions occur during the neutron decoupling~\citep{Bahcall:2000sa}. The $np$ optical depth is around unity at the decoupling radius by definition, and the energy of quasithermal neutrinos is~\citep{Bahcall:2000sa}
\begin{equation}
E_\nu^{\rm qt} \approx0.1\Gamma_{n,\rm dec} m_pc^2/(1+z),
\end{equation}
which predicts $\sim1-10$~GeV neutrinos. 

The neutrino energy fluence is estimated to be
\begin{eqnarray}
E_\nu^2 \phi_{\nu_\mu}&\approx&\frac{1}{12}\frac{(1+z)}{4 \pi d_L^2}\zeta_n\left(\frac{\Gamma_{n,\rm dec}}{\Gamma}\right)\xi_N{\mathcal E}_\gamma^{\rm iso}\nonumber\\
&\sim&0.01~{\rm erg}~{\rm cm}^{-2}~\zeta_n(\Gamma_{n,\rm dec}/0.2\Gamma)\xi_{N,1}{\mathcal E}_{\gamma,54.5}^{\rm iso},\,\,\,\,\,\,\,
\end{eqnarray}
where a nucleon inelasticity of $\approx0.5$ in $np$ collisions is taken into account, and the other $1/6$ comes from the fact that $2/3$ of pions produced by $np$ collisions are charged pions and $3/4$ of their decay products are equally shared by each flavor of neutrinos after the neutrino mixing\footnote{This is a good approximation especially for successful jets but neutrinos produced deep inside a star are subject to neutrino oscillation due to effects of matter and neutrino self-interaction~\citep{Carpio:2020app,Abbar:2022hgh}.} 
Also, $\zeta_n$ is the number ratio of neutrons to protons, $\xi_N{\mathcal E}_\gamma^{\rm iso}$ is the kinetic energy of the proton outflow with $\Gamma\gtrsim\Gamma_{n,\rm dec}$ and $\xi_N$ is the nucleon loading factor. 

In Figure~\ref{fig:qtnu} left, we show neutrino fluences in the neutron ``decoupling" model with $\zeta_n=1$. We set $\Gamma_{n,\rm dec}$ from Eq.~(\ref{eq:ndec}) assuming $\Gamma=\Gamma_{\rm max}$. The spectra of neutrinos from $np$ collisions are calculated with \texttt{Geant4} following \cite{Murase:2013hh}.

\subsection{Neutrinos from colliding neutron-loaded flows} 
If the neutron decoupling occurs before $\Gamma\approx\Gamma_{\rm max}$ is achieved, the neutron flow will be caught up by the proton flow, leading to $pn$ collisions~\citep{Beloborodov:2009be,Meszaros:2011hr}. Alternatively, if the coasting occurs earlier than the decoupling, the dissipation of neutrons via internal collisions between the compound flows may happen. Such collisions are expected around $r_{\rm dec}\ll r_{\rm ph}$, where the $pn$ optical depth is $\tau_{pn}\approx1~(\Gamma/\Gamma_{n,\rm dec})(\zeta_n/\zeta_e)\tau_{T}$. 
The typical energy of neutrinos is given by~\citep{Murase:2013hh} 
\begin{equation}
E_\nu^{\rm qt} \approx0.1\Gamma\Gamma_{\rm rel}^\prime m_pc^2/(1+z),
\end{equation}
where $\Gamma_{\rm rel}^\prime\sim2$ is the relative Lorentz factor of the interacting flow and $\sim30\mbox{--}300$~GeV neutrinos are expected for $\Gamma\sim{10}^2\mbox{--}{10}^3$. 

The neutrino energy fluence is
\begin{eqnarray}
E_\nu^2 \phi_{\nu_\mu}&\approx&\frac{1}{12}\frac{(1+z)}{4 \pi d_L^2}\tau_{pn}\xi_N{\mathcal E}_\gamma^{\rm iso}\nonumber\\
&\sim&0.03~{\rm erg}~{\rm cm}^{-2}~\tau_{pn}\xi_{N,1}{\mathcal E}_{\gamma,54.5}^{\rm iso}
\end{eqnarray}
where the normalization is set by $\xi_N{\mathcal E}_\gamma^{\rm iso}$ as the kinetic energy of the interacting flow. 
It has been suggested that dissipation induced by internal collisions between neutron-loaded flows may be relevant for subphotospheric dissipation~\citep{Beloborodov:2009be,Meszaros:2011hr}, in which $E_\nu^2 \phi_\nu\sim E_\gamma^2 \phi_\gamma$ and $\xi_{N}\sim3-30$ can be considered as fiducial values. 

In Figure~\ref{fig:qtnu} left, we show neutrino fluences in the ``collision'' model for $\Gamma=300$ and $\Gamma=800$, where $\tau_{pn}=1$ is assumed.  

\subsection{Implications}
We calculate the number of signal events using the latest all-flavor effective areas for GRECO selection~\citep{IceCube:2020qls,IceCube:2023rhf} and through-going muon neutrinos~\cite{IceCube:2021xar} in IceCube\footnote{Note that in the arXiv version the results presented in Figure~\ref{fig:qtnu} are updated using the new GRECO effective area provided by \cite{IceCube:2023rhf}.}. The values of ${\mathcal N}_{\rm sig}$ for different models are shown in Figure~\ref{fig:qtnu} left. 
Although the decoupling model is difficult to test with IceCube and other detectors such as KM3Net and Baikal-GVD, we find that the collision model is promising. A few events of $\sim100$~GeV neutrinos can be detected for $\xi_N\sim10$ especially if the Lorentz factor is sufficiently large (e.g., $\Gamma\sim800$). These results are encouraging and we urge dedicated searches for GeV--TeV neutrinos for GRB 221009A with the existing IceCube data. 

In Figure~\ref{fig:qtnu} right, we also show the sensitivity to $\xi_{N}$ as a function of $\Gamma$ for double and triplet detections of signal neutrinos from GRB 221009A. Although the expected signal can dominate if angular uncertainty is not far from the kinematic angle~\citep{Murase:2013hh}, the actual detectability depends on the atmospheric background rate, so dedicated analyses at sub-TeV energies are necessary. A search time window ($\Delta T$) will also need to be carefully considered. The burst duration may vary depending on energy bands~\citep[e.g.,][]{Zhang:2013coa}, and the engine duration is uncertain. Neutrino emission may be dominated by the main episode that lasts for $\Delta T\sim100$~s, and luminosity-weighted searches could also be helpful in more general. 

Note that in both the decoupling and collision models, neutrinos and gamma-rays are mainly produced deep inside the photosphere, from which gamma-rays with $\varepsilon_\gamma\gtrsim\Gamma m_ec^2$ cannot escape~\citep[e.g.,][]{Murase:2007ya}.
Residual neutrons would eventually decay after $\sim880/\Gamma_{n,\rm dec}$~s in the observer frame, but the resulting electron antineutrino energy is $\sim0.48~\Gamma_{n,\rm dec}$~MeV, which is difficult to detect with IceCube-like detectors. Electrons may lose their energies via synchrotron and inverse-Compton emission but their signatures may easily be overwhelmed by other components.   

%Finally decoupling neutrons have $n\rightarrow$, which gives 30~MeV antielectron neutrinos and electrons. 
%They will decay after the prompt phase and produce X-rays by upscattering optical afterglow photons. 

\section{Summary and Discussion}\label{sec3}
We considered how observations of neutrinos from the brightest GRB can be used for constraining GRB model parameters, including the CR baryon loading factor that is a critical parameter for the production of high-energy neutrinos and UHE CRs. 
We showed that the IceCube non-detection of TeV-PeV neutrinos from GRB 221009A leads to intriguing constraints on the parameter space of $r_{\rm diss}$, $\Gamma$, and $\xi_{\rm cr}$, which are comparable to those from the stacking analysis that is subject to systematics from many GRBs with different properties. We found that CR acceleration near the photosphere is likely to be subdominant and obtained $\xi_{\rm cr}\lesssim1$ for $\Gamma\lesssim300$. 
We also pointed out that the non-detection of high-energy nonthermal neutrinos is not surprising in light of the gamma-ray constraint. This is consistent with outer-zone models~\citep[e.g.,][]{Murase:2008mr,Zhang:2012qy}. However, neutrinos and gamma-rays may come from different dissipation regions, and further investigation with multi-zone models \citep{Bustamante:2016wpu,Rudolph:2019ccl} might be relevant. 

Quasithermal neutrinos, which are naturally expected if neutrons are loaded into GRB outflows, are not yet constrained. We found that in the collision model the detection of GeV-TeV neutrinos is possible with IceCube's low- and high-energy channels, or reasonable constraints on $\xi_N$ can be obtained. Even higher-energy neutrinos may be produced via the neutron-proton-converter acceleration mechanism~\citep{Kashiyama:2013ata}, and we encourage dedicated searches by considering appropriate time windows focusing on the bright episodes of prompt emission.

Finally, we note that UHE CRs could be accelerated by external shocks during the early afterglow phase~\citep{Waxman:1999ai,Murase:2007yt,Razzaque:2013dsa}, in which PeV-EeV neutrinos are expected and the predicted fluxes have not been reached by the current IceCube. Future UHE neutrino detectors~\citep{Ackermann:2022rqc} such as IceCube-Gen2, Trinity, and GRAND will be required to test those afterglow models.

%%%%%%%%%%%%%%%%%%%%%%%%%%%%%%%%%%%%%%%%%%%%%%%%%%
%%%%%%%%%%%%%%%%%%%%%%%%%%%%%%%%%%%%%%%%%%%%%%%%%%

\begin{acknowledgements}
We thank Francis Halzen and Kazumi Kashiyama for discussions. The work of K.M. and M.M. is supported by the NSF Grant No.~AST-2108466. 
K.M. also acknowledges support from the NSF Grants No.~AST-1908689, No.~AST-2108467, and KAKENHI No.~20H01901 and No.~20H05852. M.M. also acknowledges support from the Institute for Gravitation and the Cosmos (IGC) Postdoctoral Fellowship. A.K. acknowledges support from Nevada Center for Astrophysics.
S.S.K. acknowledges support by the Tohoku Initiative for Fostering Global Researchers for Interdisciplinary Sciences (TI-FRIS) of MEXTs Strategic Professional Development Program for Young Researchers. The work of K.F is supported by the Office of the Vice Chancellor for Research and Graduate Education at the University of Wisconsin-Madison with funding from the Wisconsin Alumni Research Foundation. K.F. acknowledges support from NASA (NNH19ZDA001N-Fermi, NNH20ZDA001N-Fermi) and National Science Foundation (PHY-2110821).  

%{\it Note added.}
While we were finalizing the manuscript, the related work~\citep{Ai:2022kvd} came out. The approaches are different, and we impose constraints on GRB parameters based on projecting expected events with model spectra, instead of the reported single power-law upper limit. Our study also includes calculations of nonthermal as well as quasithermal neutrinos.
\end{acknowledgements}

\bibliographystyle{aasjournal}
\bibliography{kmurase}

%\appendix
\end{document}